\documentclass[aps,preprint]{revtex4-1}%
\usepackage{amssymb}
\usepackage{amsfonts}
\usepackage{amsmath}
\usepackage{graphicx}%
\providecommand{\U}[1]{\protect\rule{.1in}{.1in}}

\begin{document}
\preprint{HEP/123-qed}
\title[The effect of local thermal fluctuations on the folding...]{The effect of local thermal fluctuations on the folding kinetics: a study from
the perspective of the nonextensive statistical mechanics}
\author{J. P. Dal Molin}
\affiliation{Departamento de F\'{\i}sica e Qu\'{\i}mica, FCFRP, Universidade de S\~{a}o
Paulo, 14040-903 Ribeir\~{a}o Preto, SP, Brazil}
\author{M. A. A. da Silva}
\affiliation{Departamento de F\'{\i}sica e Qu\'{\i}mica, FCFRP, Universidade de S\~{a}o
Paulo, 14040-903 Ribeir\~{a}o Preto, SP, Brazil}
\author{A. Caliri}
\affiliation{Departamento de F\'{\i}sica e Qu\'{\i}mica, FCFRP, Universidade de S\~{a}o
Paulo, 14040-903 Ribeir\~{a}o Preto, SP, Brazil}
\keywords{one two three}
\pacs{PACS number}

\begin{abstract}
Protein folding is a universal process, very fast and accurate, which works
consistently (as it should be) in a wide range of physiological conditions.
The present work is based on three premises, namely: ($i$) folding reaction is
a process with two consecutive and independent stages, namely the search
mechanism and the overall productive stabilization; ($ii$) the folding
kinetics results from a mechanism as fast as can be; and ($iii$) at nanoscale
dimensions, local thermal fluctuations may have important role on the folding kinetics.

Here the first stage of folding process (search mechanism) is focused
exclusively. The effects and consequences of local thermal fluctuations on the
configurational kinetics, treated here in the context of non extensive
statistical mechanics, is analyzed in detail through the dependence of the
characteristic time of folding ($\tau$) on the temperature $T$ and on the
nonextensive parameter $q$.

The model used consists of effective residues forming a chain of 27 beads,
which occupy different sites of a $3-$D infinite lattice, representing a
single protein chain in solution. The configurational evolution, treated by
Monte Carlo simulation, is driven mainly by the change in free energy of
transfer between consecutive configurations.

We found that the kinetics of the search mechanism, at temperature $T$, can be
equally reproduced either if configurations are relatively weighted by means
of the generalized Boltzmann factor ($q>1$), or by the conventional Boltzmann
factor ($q=1$), but in latter case with temperatures $T^{\prime}>T.$

However, it is also argued that the two approaches are not equivalent. Indeed,
as\ the temperature is a critical factor for biological systems, the folding
process must be optmized at a relatively small range of temperature for the
set of all proteins of a given organism. That is, the problem is not longer a
simple matter of renormalization of parameters. Therefore, local thermal
fluctuation on systems with nanometric components, as proteins in solution,
becomes a important factor affecting the configurational kinetics.

As a final remark, it is argued that for a heterogeneous system with
nanoscopic components, $q$ should be treated as a variable instead of a fixed parameter.

\end{abstract}
\volumeyear{year}
\volumenumber{number}
\issuenumber{number}
\eid{identifier}
\date[Date text]{date}
\received[Received text]{date}

\revised[Revised text]{date}

\accepted[Accepted text]{date}

\published[Published text]{date}

\startpage{101}
\endpage{102}
\maketitle

\section{Introduction}

Differently of most polymers, each natural protein folds over itself in a
specific $3-$D structural conformation, its native structure. The series of
events that drive a polypeptidic chain into its native structure, the folding
process, is not yet fully understood: protein systems involve many complex
interactions, and presents several remarkable properties that seems to require
new experiments \cite{kumar}, theoretical, and computational approaches
\cite{dalmolin, mok, dill}.

Although the folding process is surprisingly quick, folding rates $(K_{f})$ of
different proteins can span several orders of magnitude ($K_{f}$ is a measure
of how fast the folding process leads the chain from the unfolded state up to
the native structure), and this remains true even for proteins of
approximately the same size. Moreover, even for a single domain, two states,
small proteins, existing theories for the kinetics of folding can not
quantitatively predict this experimental observation \cite{plaxco} --probably
because the folding mechanisms have been routinely proposed from their
ensemble-averaged properties, and from conflicting interpretation of its
fundamentals \cite{rose1, rose2}. \ Hence, the alternative ideas and
hypotheses about folding explain only partially the phenomenon. For instance,
the concept of transition state explain satisfactorily the two state kinetics
but not the folding reaction rates, while that the funnel\ landscape idea can
give insights about the folding rates but not for the two state kinetics
\cite{Dill-a}. \ 

A remarkable characteristic of the folding process is its robustness, which
can be illustrated by two intriguing properties: first, one finds that the
folding is similarly processed in a large temperature range, covering about
100$\,^{\mathtt{o}}$C; and second, all functional proteins of all organisms,
which live in the most different environments, fold correctly, and are stable
about a particular ideal temperature. Indeed, living organisms are found in
extreme conditions: some live in environments with temperatures near to
freezing water \cite{piette, Nicholas}, while others are found in places with
temperatures of boiling water \cite{wiggins, zebb}. Therefore, the search
mechanism must work properly in the temperature interval from about zero to
about 100 $^{\mathtt{o}}$C, while that for each alive species the range of
functional temperature is in general relatively much smaller.

Proteins also present an extraordinarily precise and fast self-organization
process. They fold some ten orders of magnitude faster than the predicted rate
of a random search mechanism \cite{levinthal}; it is as if each protein had
been designed to fold as fast as possible. Indeed, the probability of finding
a fast-folding sequence, choosing it randomly from the set of all possible
sequences, is very small \cite{gutin}. However, there is also a physiological
reason for fast folding: because do not have enough chaperone molecules to
support folding of every protein (anyway chaperones also are proteins), they
must fold very rapidly in order to avoid aggregation due to exposing
hydrophobic areas of their surface for too long \cite{lorimer, ecroyd}.

Globular proteins can be considered as independent nanomachines. This
particularity, in combination with the nature of most currently available
experimental data, can be considered as one of the sources of certain
inadequate views of the folding problem. The stable appearance and the
properties of homogeneous macroscopic objects are resulting from the average
activity of a very large number of atoms, but, contrasting with this scenario,
nanostructures like colloidal particles or proteins, in contact with a thermal
reservoir (the solvent), experience thermal fluctuations in a special way
\cite{beck, rajagopal}. Actually, local unbalanced forces shake and deform
continuously each of such nanostrucutures, which cannot be revealed by most of
available data about protein kinetics that just reflects the collective
behavior of a huge number of them in dilute aqueous solutions. That is, the
result is a kind of temporal averaged view of the phenomenon. Nevertheless,
new data and ideas start to emerge from single molecule experiments
\cite{kumar}, such as about transition paths at equilibrium --which is only
observable for single molecules, allowing to obtain crucial mechanistic
information, for instance, folding and unfolding rates \cite{Eaton}.

Part of such general properties may be better understood if one considers that
the search mechanism is governed mainly by the hydrophobic effect, whose
strength, as shown experimentally at least for small hydrophobic molecules
\cite{chandler, luiz}, varies slightly in the temperature interval from about
zero to 100 $^{\mathtt{o}}$C. Therefore, it is suggested that folding process
should be composed of two temporal steps \cite{eulalia}: the search mechanism,
as the first stage, followed by the overall stabilization that only begins
with the chain close enough to its native conformation, when energy and
structural requirements, as encoded in the residue sequence, would be
associated in a productive and cooperative way.

Based in these general properties, a few hypotheses can be formulated;
therefore we assume as general grounds for the folding problem, the following
three statements:

$(i)$- the complete folding process is composed by two timely independent
steps, namely: the search mechanism, and the overall productive stabilization.

$(ii)$- for typical one domain, two state globular proteins the folding
instructions encoded in the residues sequence provide a folding kinetic as
fast as possible; and

$(iii)$- at nanoscale dimensions randomness emerges as a peculiar attribute of
the protein molecule, which should be treat individual and appropriately with
respect to the effects of local thermal fluctuations.

Our goal in this work is to show evidences concerning the importance of local
thermal fluctuations on the kinetics of the folding process of globular
proteins. The simplified model employed here (next section) focuses
exclusively on the search mechanism and have as its grounds the hydrophobic
effect. Fluctuation effects on a nanoscale structure are treated in the
context of the nonextensive statistical mechanics (section III), and are
analyzed in details through the dependence of the folding characteristic time
$\tau$ on the temperature $T$ and nonextensive parameter $q$ (section IV). The
behavior of small, single-domain globular proteins are used here as ideal
prototypes; usually many of them fold via an all-or-nothing process, that is,
without detectable intermediates \cite{jackson}. Comments and conclusions
(section V) are formulated according the three hypotheses stated above.

\section{The model}

The model presented here is based on the first hypothesis, stated in previous
section. It is devoted just for the first stage of the folding process, the
search mechanism, in order to explore general aspects of the folding problem
valid, in principle, for all proteins. Therefore a lattice model is used:
effective residues (a chain of 27 beads), occupying consecutive and distinct
sites of a three-dimensional infinity cubic lattice, represent a single
protein-like chain in solution; effective solvent molecules, which explicitly
interact with the chain, fill up the lattice vacant sites. The general scheme
to explore the configurational space presumes that, during the simulation,
solvent molecules and chain units exchange their respective sites so that all
sites of the lattice remain always fully filled \cite{eulalia, roosevelt}.

For each configurational change, only the transfer free energy (variations on
the hydrophobic energy) is taken into account --given that the model is
conceived to deal specifically with the search mechanism; solvent-solvent and
residue-residue interactions are represented by hard core-type interactions
(excluded volume). For a regular cubic lattice, which in the present case
means uniform solvent density, this interaction scheme is exactly equivalent
to the use of additive, first neighbor, inter-residue pairwise potentials,
namely $h_{i,j}=h_{i}+h_{j}$, where $h_{i}$ is the hydrophobic level of the
$i^{th}$ residue in the chain sequence \cite{luiz2}. Residues are taken from a
repertory of ten distinct units (a ten-letter alphabet), which are
characterized by distinct hydrophobic levels and a set $\{c_{i,j}\}$ of
inter-residue steric specificities. The hydrophobic levels has been considered
the most general and influential chemical factor acting along the folding
process \cite{li}, while the set of inter-residues constraint mimics steric
specificities of the real residues. These specificities are achieved through
the specification of which pairs of residues are allowed to get closer, as
first neighbors, and its main consequence is to select folding and unfolding
pathways through the configurational space. The set of inter-monomer
constraints is fixed for each monomer pair, that is, it does not depend on the
particularities of the native structure \cite{dalmolin, eulalia}.

The configurational energy $E([\kappa,l])$ of an arbitrary chain configuration
$\xi$ defined by the set $[\kappa,l]$ of $N_{\xi}$ first neighbor
inter-residue contacts $(0\leq N_{\xi}\leq28)$ is%

\begin{equation}
E([\kappa,l])=\sum_{\{i,j\}}(h_{i,j}+c_{i,j})\delta_{(i,j),[\kappa,l]},
\label{0}%
\end{equation}
where the sum runs over the set of all residues pairs $\{i,j\}$; the factor
$\delta_{(i,j),[\kappa,l]}=1$ if $(i,j)$ belongs to the set $[\kappa,l]$;
otherwise $\delta_{(i,j),[\kappa,l]}=0$.

The present model is not native-centric, that is, data from the native
structure are not employed to guide the chain along the simulation.
Consequently, a rule for sequence designing, valid for any target structure
--representing the native structure, is necessary. The provided syntax is
mainly based on the \textit{hydrophobic inside} rule \cite{dill2} and on the
local topological features of the target structure \cite{roosevelt, tese}.

Models based on this stereochemical potential has been proved to be efficient
in packing the chain and finding the native state \cite{ines}, but they fail
to provide stability to the native state because such additive potential,
$h_{i,j}=h_{i}+h_{j}$, satisfies marginally the segregation principle (namely,
$2h_{i,j}+h_{i,i}+h_{j,j}\leq0$) through the equal sign, that is:
$2h_{i,j}+h_{i,i}+h_{j,j}=0$. However, adding up steric constraints
$\{c_{i,j}\}$ to the hydrophobic potential $h_{i,j}$, as in Eq.(\ref{0}), some
important consequences are observed: for instance, it helps to select folding
and unfolding pathways, which makes faster the folding process, and improves
the overall stability condition of the globule in the native state
\cite{eulalia}.

The folding process is simulated through the Metropolis Monte Carlo (MC)
method, involving standard elementary chain moves, namely: crankshaft, corner,
and end flips. For each move attempt, a particular reference point along the
chain is chosen at random and the process evolves without any reference to the
native configuration, except to check when it is found for the first time: for
each particular run, the number of MC-steps spent to reach the native
structure from the initial configuration (the first passage time) is took as
the folding time $t$ for that case.

\section{Local thermal fluctuations and the nonextensive statistical
mechanics}

Usual thermodynamic and kinetic data about proteins are time-averaged results
from the collective behavior of many molecules, something between$~\sim
10^{17}-10^{20}$ molecules/litter. This condition determines the traditional
tendency to view globular proteins as mostly compact and static structures.
However, when considered individually, proteins surely undergo strong
fluctuations in their thermodynamic properties. For instance, let us reproduce
here a specific thermodynamic calculation \cite{cooper} for a system
constituted by a representative protein of about $250$ residues in solution,
\ with molecular mass $m$ about $4\times10^{-23}$ kg that typically shows heat
capacity $(C_{p}\cong C_{v})$ about $0.3$ kcal kg$^{-1}$K$^{-1}$, at
temperature $T=300$K. A simple estimate of the internal energy fluctuation
about the mean, for an individual molecule, gives $\Delta U_{rms}\cong
6\times10^{-20}$ cal per molecule $(\Delta U_{rms}=kT^{2}mC_{V}) $.
Essentiality, it would be in the same order of magnitude of the typical
enthalpy changes on thermal denaturation of proteins --tens of Kcal/mol
\cite{privalov}-- if all molecules were fluctuating in concert \cite{cooper}.
Actually, fluctuations are individually uncorrelated: in a population with a
huge number of macromolecules, fluctuations tend to cancel each other,
producing thermodynamic parameters well behaved.

However, for each single protein such fluctuations may interfere in the
folding kinetics in one way or another; therefore the folding process has to
be explained for each single molecule. Let us then think about a protein in
solution as a heterogeneous system constituted by just one chain in its
solvent, which works as a heat reservoir at macroscopic temperature $\beta
_{0}^{-1}$. But, due to the protein nanosize scale, it is as if the
temperature were locally fluctuating. Then, if (locally) $\beta^{-1}$
fluctuates rapidly with respect to the typical time spent for chain
configurational interchanges, one could think about a generalized Boltzmann
factor exp$_{q}(-\beta_{0}\epsilon)$ as an integral over all possible locally
fluctuating $\beta^{-1}$, that is%

\begin{equation}
\mathrm{exp}_{q}(-\beta_{0}\epsilon)=\int_{0}^{\infty}\mathrm{exp}%
(-\beta\epsilon)f(\beta)d\beta. \label{1}%
\end{equation}
It has been shown that if $f(\beta)$ is assumed to be the $\chi^{2}%
$-distribution, a special case of the gamma-distribution of variable $\beta$,
present in many common circumstances \cite{hastings}, the generalized
Boltzmann factor becomes \cite{beck, rajagopal, touchette}%

\begin{equation}
\mathrm{exp}_{q}(-\beta_{0}\epsilon)=[1-(1-q)\beta_{0}\epsilon]^{\frac{1}%
{1-q}}, \label{2}%
\end{equation}
which is the same expression proposed in the context of nonextensive
statistical mechanics \cite{tsallis1, tsallis2}. The $\chi^{2}$-distribution%

\begin{equation}
f(\beta)=[\Gamma(n/2)]^{-1}\left(  \frac{n}{2\beta_{0}}\right)  ^{n/2}%
\beta^{-1+n/2}\exp\left(  -\frac{\beta}{2\beta_{0}}n\right)  \label{3}%
\end{equation}
is parameterized such that the heat reservoir temperature $\beta_{0}^{-1}$
coincides with the average of the fluctuating $\beta$, that is: $\beta
_{0}=\langle\beta\rangle=\int_{0}^{\infty}\beta f(\beta)d\beta$ . The
nonextensive parameter $q$, set as%

\begin{equation}
q=1+2/n, \label{4}%
\end{equation}
is associated with the relative dispersion of $\beta$, according
$q=1+(\langle\beta^{2}\rangle-\beta_{0}^{2})/\beta_{0}^{2}$, where $n$ is the
number of degrees of freedom. This point will be returned later in the next section.

In the present work our interest is mainly concerned with the kinetic behavior
of the chain during the folding process, starting from a open configuration
until to reach its native conformation. Therefore, for MC realizations we
assume a generalized transition probability exp$_{q}(-\beta_{0}\Delta
\epsilon_{ab})=\langle$exp$(-\beta_{0}\Delta\epsilon_{ab})\rangle$ between the
configuration $a$ with energy $\epsilon_{a}$, and configuration $b$ with
energy $\epsilon_{b}$, that is,%

\begin{equation}
\mathrm{exp}_{q}(-\beta_{0}\Delta\epsilon_{ab})=[1-(1-q)\beta_{0}%
\Delta\epsilon_{ab}]^{\frac{1}{1-q}} \label{5}%
\end{equation}
where $\Delta\epsilon_{ab}=\epsilon_{b}-\epsilon_{a}$ .

For $q\gtrsim1$ the $q$-exponential and the conventional exponential function
behave in a very similar way, that is, one may expect that exp$_{q}(-x)$ is
effectively equivalent to the conventional exponential function in which its
argument has been adequately changed (that is, with the temperature increased
somewhat). Actually, comparing the two following difference functions, namely
$\Delta_{q}(x)=\exp_{q}(-x)-\exp(-x)$ and $\Delta_{a}(x)=\exp(-ax)-\exp(-x)$,
as shown in Figure \ref{figure1} for $a $ and $q\cong1$ , one sees that their
profiles are similar, although the maximum of $\Delta_{a}(x)$ occurs about
$x_{a}=1,$ and for $\Delta_{q}(x)$ it is about $x_{q}=2$. Actually, the
$\lim_{a\rightarrow1}(x_{a})=1,$ while $\lim_{q\rightarrow1}(x_{q})=2$; in
both cases $x_{a}$ and $x_{q}$ increases slowly when $a$ decreases from one
and $q$ increases from one.

Therefore we compare the effect of both approaches on the configurational
kinetics through MC simulation, and then discuss possible implications for the
understanding of the folding mechanism. In order to address directly this
issue, we consider the folding time $t$ and the folding characteristic time
$\tau$ (described in the next section) as analytical amounts emerging from the
folding kinetics. The comparison between the two approaches emphasizes the
effects of local fluctuations in a heterogeneous system (with nanosized scaled
components) through the view of the nonextensive statistical mechanics.

\section{Results and Discussion}

The characteristic folding time $\tau$ is determined by means of a sample of
many folding trajectories, that is for each target (native) structure a number
$N$ of independent runs represents the folding process of a set of $N$
non-interacting proteins (diluted solution). For each run, say the $i^{th}$
run, the MC time $t_{i}$ spent to find the native structure (first passage
time) is adopted as the folding time for that case. So, at the end of $N$
independent runs one gets a set $\{t_{i}\}_{N}$ of independent folding times
and then, by counting the number of folding times that fall in each time
interval $[t,t+\Delta t]$ one finally gets the decay histogram of the number
of unfolded proteins as a function of the MC time $t$. These data are then
fitted by one (or more) exponential function, giving the specific
characteristic folding time $\tau$ for that structure. The simulations are
carried out in a given range of temperatures $T$ for several values of the
nonextensive parameter $q$, and for distinct native structures. Each native
structure is characterized by their topological complexities, which can
roughly be estimated by its structural Contact Order \cite{plaxco1, ines}.

As an encompassing survey, Table $1$ shows $\tau$ in function of the
temperature $T$ and nonextensive parameter $q$, in the interval $0.9\leq
T\leq1.5$ and $1\leq q\leq1.4$. Three representative target (native)
structures (identified as ID 866; 1128 and 36335) were used; in general,
$\tau$ depends on the structure complexity, and is a continuous, convex
function of $T$ and $q$. A total of $N=150$ independent runs were used for
each pair $(T,q)$. The structure ID 36335 presents higher topological
complexity than the others two, a fact reflected in its larger $\tau$.

For any temperature $T_{i}$, there is a specific $q_{i}=q(T_{i}),$ let us say
$q_{i}^{\ast},$ that minimizes $\tau$, that is $\tau\rightarrow\tau_{\min}$,
as emphasized in Table $1$ by shaded cells; better approximations can be
achieved by extra refinement of $q$. \ The uncertainty in $\tau$ was estimated
by the standard deviation of the mean of means, considering $20$ distinct
samples $\{t_{i}\}_{N}$ $($with $N=150)$ taken from an extended set of
$10^{4}$ independent runs. The uncertainty $\delta\tau$ depends on the pair
$(T,q);$ the smallest uncertainties $(\simeq5\%)$ occur for those specific
values $q_{i}^{\ast}$ which minimize the characteristic folding time
$\tau\cong\tau_{min}$. On the other hand, when the system approaches the
glassy regime ($T<<1)$ the time spent in metastable states increases
substantially, and so $\delta\tau$ is strongly influenced by the size $N$ of
the set $\{t_{i}\}_{N}$ of independent runs.

This scenario suggests that the kinetic of the search mechanism is equally
reproduced, not mattering if the configurations are relatively weighted by
mean of the generalized Boltzmann factor, namely, $\exp_{q}(-\Delta
\epsilon/kT)$ (see Eq. (\ref{2})), or by the conventional Boltzmann factor,
$\exp(-\Delta\epsilon/kT^{\prime}~),$ with the system temperature $T^{\prime}$
increased by some amount with respect to $T$. This can be seen clearly if, for
each $q,$ the behavior of $\tau$ is plotted as a function of the translated
temperature scale $\mathcal{T}_{q}=T-T_{q}^{\ast}$, as shown in Figure
\ref{figure2} for structure ID 1128; $T_{q}^{\ast}$ is the temperature in that
$\tau$ approaches $\tau_{\min}$ for that value of $q$. Essentially all curves
behave in the same way about $\mathcal{T}_{q}=0.$

A more detailed examination shows that the distributions $\Phi$ of folding
times are essentially the same in both approaches, that is, using $\exp
_{q}(-\Delta\epsilon/kT)$ \ or $\exp(-\Delta\epsilon/kT^{\prime}).$ Figure
\ref{figure3}\ \ shows $\Phi=\Phi(T)$ for structure ID $866,$ for different
values of $q$; much more\ independent runs $(N=10^{4})$ were employed for each
case. In general, the distributions are better fitted with one or more
lognormal curves, depending on the temperature.

For $(q,T)=(1,1)$, the system approaches the glassy regime with manifestation
of ergodic difficulties, as indicated by the three peaked curve; Figure
\ref{figure3}, open, smaller circles. As $T$ increases from $T=1,$ the domain
of $t$ is accordingly reduced; at $T\cong1.5$ the distribution presents the
smallest domain, namely $0<\ln(t)<5.5$, and as $T$ increases from this point
its behavior is reverted: the size of the domain starts to increase again and
the curve's peak moves in the direction of larger $t$. In the region of the
smaller folding times, namely for $\ln(t)<2.5$, all curves present the same
behavior, if $T$ \ is restricted in the interval $1.2\leq T\leq1.5$. The
meaning for this is: even with the temperature 25\% higher than $T=1.2$, there
exist some configurations among the initial open ones, which combined with
certain configurational evolution, can lead the chain very rapidly into the
native structure. Note that this is also true for the case $(q,T)=(1.1,1.0).$

The frequency distribution $\Phi=\Phi(T)$ for $(q,T)=(1.1,1.0)$ (open large
circles --generalized Boltzmann factor) is practically the same as that for
$(q,T)=(1.0,1.25)$ (full smaller circles --conventional Boltzmann factor ),
implying in the convergence of the folding characteristic time for the two
cases, namely, $\tau_{\min}=24$ (see Table $1$ and Figure \ref{figure3}). This
result confirms that (for the present problem) the net effect of the
generalized Boltzmann weight on the kinetic of the search mechanism is
equivalent, from the perspective of the conventional Boltzmann factor, to a
specified increase in the system temperature, that is, a certain increase on
thermal fluctuations. So, a specific well tuned amount of thermal fluctuation
is what determines the fastest folding process. Then, independent of the
approach (generalized or conventional Boltzmann factor), and according the
hypothesis that the folding instruction encoded in the residue sequence
provides a folding kinetic as fast as possible (Section II, second premise),
$\tau_{\min}$ is adopted as the optimum $\tau,$ the actual characteristic
folding time.

However, due mostly to the peculiarities of protein systems, the folding
process must be minimaly optimazed in a relatively narrow range of
temperature, and for the total set of proteins of each living organism. \ In
this sense the two approaches are not equivalent: the fact that each target
structure has a proper temperature for fastest folding could be seem as a
model deficience that should be improved by approaching the problem by the
nonextesive statistical mechanics. Moreover, as already mentioned, the search
mechanism operates equally in large temperature interval, but, once the native
structure is found, the other stage of the folding process takes place --the
overall productive stabilization, which is strictly dependent on the
temperature. Indeed, for the set of protein of each organism there is a
working temperature interval $T_{\omega}-\Delta T\lesssim T\lesssim T_{\omega
}+\Delta T$ \ (with $\Delta T/T_{\omega}\lesssim20\%),$ outside of which its
functionality can be seriously reduced or completely lost. Therefore, the
system temperature $T$ must be kept as the reference temperature, measured
macroscopically, and all thermal characteristics of a nanosize body, in
response to the local thermal fluctuations, should be conveniently controlled
by the nonextensive parameter $q$.

For any target structure, at specific system temperature $T$, it is always
possible to adjust $q$ in order to get the optimum $\tau$. But, what intrinsic
factors would determine a specific $q^{\ast}$ value that induces the fasted
folding for that specific protein? In the present case, namely a chain
evolving through the configurational space from a open chain into a compact
specific globule, the straightforward idea comes from the observation that
local fluctuations of $\beta$ should be dependent on the spatial scale
\cite{beck1, beck2}. Indeed, along the simulation specific traps as well wrong
packing tendency are recurrent, and so the resulting effect of local thermal
fluctuations is to promote a rich variety of shape and size of the globule. To
see this argument in more details, let each residue of the chain be
associated, as upper bound, to just one degree of freedom, which allow us to
explore the relation between $q$ and the numbers of degrees of freedom $n$ of
the system: $q=1+2/n$, Eq. (\ref{3}) and (\ref{4}). Through this relation we
may recognize a subtle association between $q^{\ast}$ and the topological
complexity of the native structure. One notes, firstly, the dynamic nature of
the number of degrees of freedom $n$: as the chain degree of compactness
changes in the course of the time, $n$ changes accordingly. So, for an open
chain we have $q_{\min}=1+2/n_{\max}\rightarrow1$, and for a fully compact
globule $q_{\max}=1+2/n_{\min}$. Using $n_{\min}=1$ as a limiting condition,
one gets a kind of upper bond for $q$, that is, $q_{\max}=3$. Actually, along
the folding process, energy and topological traps must be overcome until the
target structure is reached. Such recurrent traps keep the chain for
relatively long time in wrong conformations of different degrees of
compactness, which must be disassembled so that the folding process can be
restarted. Therefore, as the chain suffers the thermal effects differently,
depending on its compactness, the simulation process should be governed by a
variable instead of a fixed parameter $q$. However, depending on the number of
traps and their peculiarities --determined by the combination of the
complexity of the native structure and the chain sequence, a specific
$q^{\ast}$ (kind of average $q$) may be associated to each target structure;
this is what we did in this work and is showed in Table $1$ for three
different target structures. Clearly, a direct inspection of this process can
be carried out using a dynamic process that changes (appropriately) the value
of $q$ along the simulation. The implementation of this idea is now in progress.

\section{Final Comments and Conclusions}

The hypotheses about the folding reaction as a two independent stages (search
and stabilization) enabled us to place emphasis just on the search mechanism
as an universal process guided by the hydrophobic force, which performs
equally in a large range of temperatures. The premise about the fastness of
the folding process (necessary to prevent protein aggregation) was used in
order to associate the nonextensive parameter $q^{\ast}$ to each native structure.

\begin{figure}
[ptb]
\begin{center}
\includegraphics[
trim=-0.009263in 0.000000in 0.002137in 0.000000in,
height=4.843in,
width=5.047in
]%
{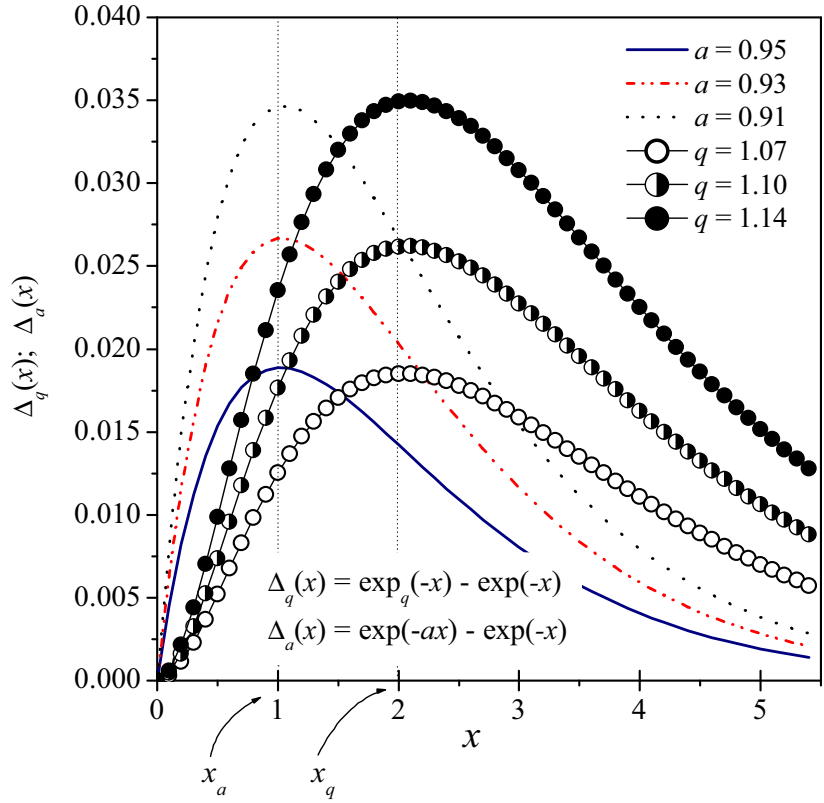}%
\caption{Functional difference between the generalized and the ordinary
exponential function, namely $\Delta_{q}(x)=\exp_{q}(-x)-\exp(-x)$, with
$q\gtrsim1,$ and between two ordinary exponential functions with the argument
of one of them rescaled, that is: $\Delta_{a}(x)=\exp(-ax)-\exp(-x),$ with
$a\lesssim1$; the parameter $a$ represents a small increment in the system
temperature. Their profiles are similar but, for $a$ and $q$ near to $1,$ the
maximum of $\Delta_{a}(x)$ and $\Delta_{q}(x)$occur about $x_{a}=1$ and
$x_{q}=2$, respectively. \ In both cases, $x_{a}$ and $x_{q}$ increases very
slowly when $a$ and $q$ depart from 1.}%
\label{figure1}%
\end{center}
\end{figure}

\begin{figure}
[ptb]
\begin{center}
\includegraphics[
height=5.2217in,
width=5.047in
]%
{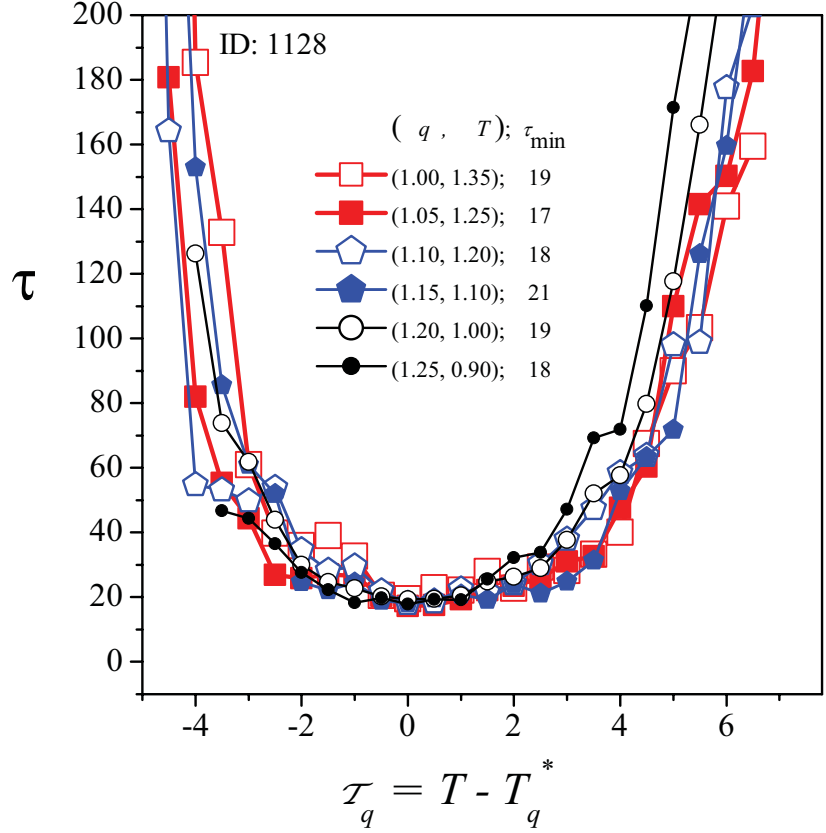}%
\caption{The $\tau\times T$ \ "U--shape " : folding characteristic time $\tau$
as a function of the translated temperature $\mathcal{T}_{q}=T-T_{q}^{\ast}$
\ for several $(q,T)$ combinations. $T_{q}^{\ast}$ is the temperature in that
$\tau=\tau_{\min}$ for that value of $q$. For each $q$, the temperature range
was covered by increments $\Delta T=0.05.$\ All curves behave very similarly
around $\mathcal{T}_{q}=0$.}%
\label{figure2}%
\end{center}
\end{figure}

\begin{figure}
[ptb]
\begin{center}
\includegraphics[
height=4.7971in,
width=5.047in
]%
{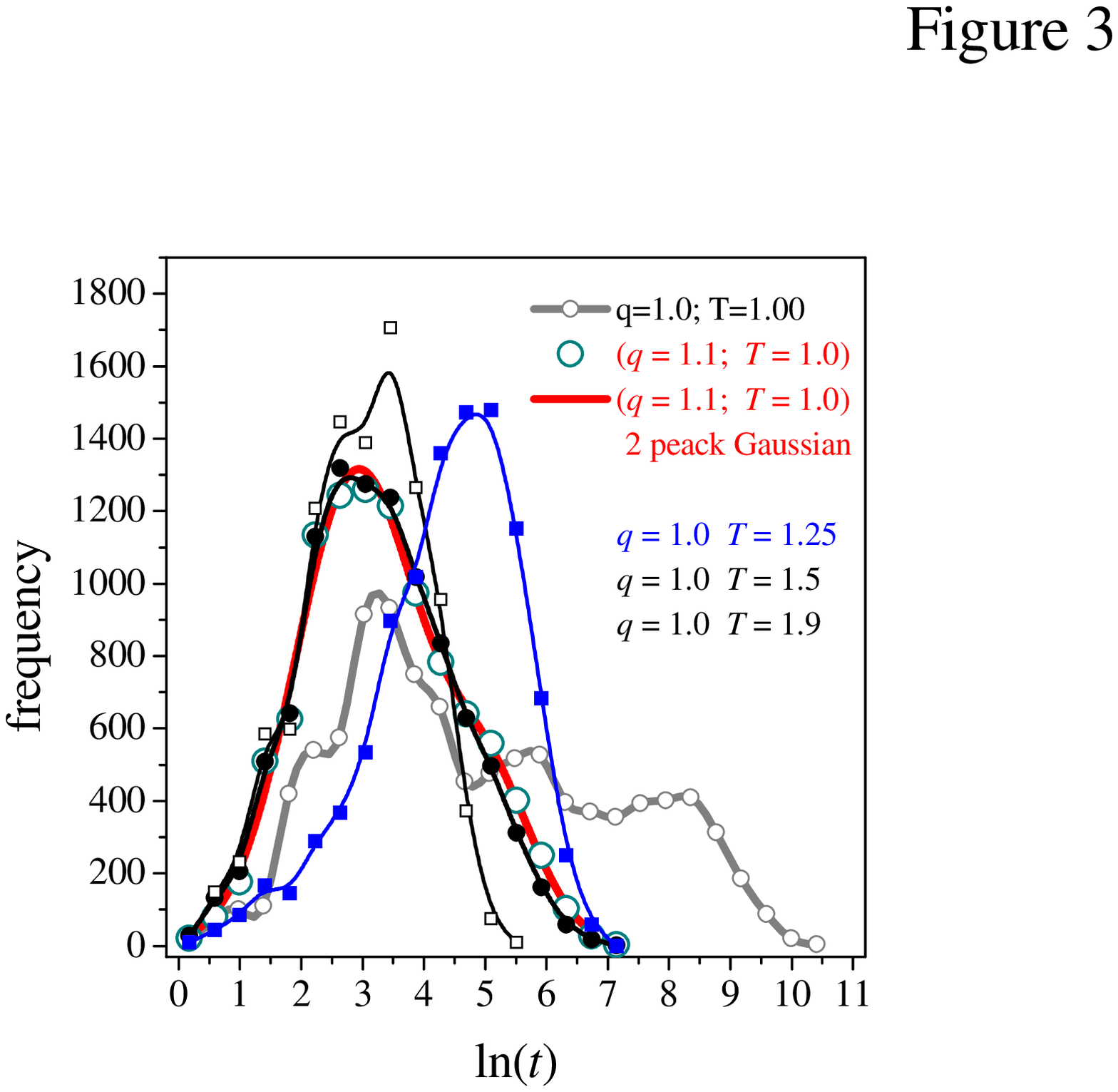}%
\caption{Folding time distributions $\Phi$ for structure ID 866. The lognormal
distribution for $q=1.1$ and $T=1.0$ \ \ (larger open circles) was fitted
using two peaks Gaussian (continuous thicker line -- amplitude: ($A_{1}%
;\,A_{2})\,=(1310\pm40;430\pm70)$; width ($w_{1};$ $w_{2})=(0.49\pm
0.07;0.8\pm0.3)$, and average $(\mu_{1};\mu_{2})=(2.91\pm0.09;5.0\pm0.2))$.
For $q=1.0$ the folding time behavior is shown for several temperatures
(fitted by basic spline functions). In particular, the folding time
distribution for $(q,T)=(1.0,1.25)$ and $(1.1,1.0)$ are practically the same,
confirming that the generalized Boltzmann factor $(q>1)$ at temperature $T$,
corresponds to the conventional Boltzmann factor $(q=1)$ with a certain
increase in $T$. In the region of the smaller folding times $(\ln(t)<2.5)$ and
for temperatures in the interval $1.0\leq T<1.5$, all curves coalesce
--including the case $(q,T)=(1.1,1.0)$. At $T=1.0$ and $q=1$ (full gray
circles) the system approaches the glassy regime and manifestation of ergodic
difficulties are evident. At temperature $T=1.5$ the distribution has the
lowest domain, that is, $0<\ln(t)<5.5$ (open squares), and inasmuch as $T$
increases from this point, the peak of the distribution moves again toward
larger $t$, as illustrated for $T=1.9$ (full squares).}%
\label{figure3}%
\end{center}
\end{figure}

The comparison between the two approaches, namely the nonextensive $(~q>1)$
and the conventional $(q=1)~$statistical mechanics, suggests that suitable
thermal fluctuations --adequately achieved only in the nonextensive context--
drives the chain through the fastest possible courses to the native
conformation, as shown in Figure \ref{figure2}. The generalized Boltzmann
factor $~$has a qualitatively equivalent effect with respect to the
conventional Boltzmann factor, that is, to enlarge the chance of removing the
chain from energetic or topological traps. Although their extremum effects on
the transition probabilities between two consecutive configurations are
energetically shifted (Figure \ref{figure1}), appropriate combinations of $T$
and $q,$ such as $(T=1;q>1)$ and $(T^{~\prime}>1;q=1),$ for example, can
produce practically the same folding time distribution $\Phi$ (Figure
\ref{figure3}), determining the same optimum characteristic folding time
$\tau_{\min}.$\ 

The well known U--shape dependence of $\tau$ on temperature, shown in Figure
\ref{figure2}, has been commonly attributed exclusively to peculiarities of
the chain sequence --or to the complexity of the target structure. Indeed,
sequences are usually generated and tested for its ability to fold rapidly in
an small and specific range of temperature \cite{Karplus-a}, even knowing that
this procedure eliminates many suitable structures that would otherwise be
important for kinetic studies. But a new perspective emerges when local
thermal fluctuations experienced by nanoscale structures is associated with
its spatial characteristics (as its size and degrees of freedom), by means of
the parameter\ $q$ \ from the nonextensive statistical
mechanics.\ Specifically, such as chaperone that assists the folding, well
tuned thermal fluctuations help to disassemble chain segments wrongly
collapsed, improving the fastness of the folding process; otherwise, using the
conventional statistic mechanics, it would be achieved only at higher
temperature of the reservoir. Therefore, extending this scenario to real
protein systems we may visualize the two main driving forces (entropic forces
compacting the chain and local thermal fluctuations tending to open it)
suporting a continuous process of folding/unfolding until, eventually, the
neighborhoods of the native state is reached. At this point, and only under
this condition, the native structural peculiarities and chain energetic
interactions, as encoded along the chain sequence, would be associated in a
cooperative and fully productive way, guarantying the globule overall stability.

As a final remark, we recall that the exploratory analysis summarized in Table
1 suggests that $q$ increases with the topological complexity of the target
structure. Indeed, treating $q$ as a variable, let us say $\mathbf{q}$, we get
essentiality the same result, that is: the characteristic time $\tau$
converges to the same $\tau_{\min}$ obtained using $q=q^{\ast}$ as a
parameter. In a preliminary investigation, $\mathbf{q}$ was functionally
linked to the instantaneous radio of gyration, which was used as a measure of
the chain compactness (degrees of freedom). Accordingly, for each of several
distinct target configurations investigated, the resulting $\mathbf{q}%
$-distribution is characterized by one or two peaks around the constant
$q=q^{\ast}$ used as a parameter.

\bigskip

\textbf{Acknowledgements} \medskip

We thank Dr. Alexandre S. Araujo for reading our manuscript, and CNPq for funding.

%

%TCIMACRO{\FRAME{ftbpFU}{5.047in}{4.843in}{0pt}{\Qcb{Functional difference
%between the generalized and the ordinary exponential function, namely
%$\Delta_{q}(x)=\exp_{q}(-x)-\exp(-x)$, with $q\gtrsim1,$ and between two
%ordinary exponential functions with the argument of one of them rescaled, that
%is: $\Delta_{a}(x)=\exp(-ax)-\exp(-x),$ with $a\lesssim1$; the parameter $a$
%represents a small increment in the system temperature. Their profiles are
%similar but, for $a$ and $q$ near to $1,$ the maximum of $\Delta_{a}(x)$ and
%$\Delta_{q}(x)$occur about $x_{a}=1$ and $x_{q}=2$, respectively. \ In both
%cases, $x_{a}$ and $x_{q}$ increases very slowly when $a$ and $q$ depart from
%1.}}{\Qlb{figure1}}{figure1.eps}{\special{ language "Scientific Word";
%type "GRAPHIC";  maintain-aspect-ratio TRUE;  display "USEDEF";
%valid_file "F";  width 5.047in;  height 4.843in;  depth 0pt;
%original-width 7.1252in;  original-height 6.9825in;  cropleft "-0.0013";
%croptop "1";  cropright "0.9997";  cropbottom "0";
%filename 'figure1.eps';file-properties "XNPEU";}}}%
%BeginExpansion
%EndExpansion

\bigskip%

%TCIMACRO{\FRAME{ftbpFU}{5.047in}{5.2217in}{0pt}{\Qcb{The $\tau\times T$
%\ "U--shape " : folding characteristic time $\tau$ as a function of the
%translated temperature $\mathcal{T}_{q}=T-T_{q}^{\ast}$ \ for several $(q,T)$
%combinations. $T_{q}^{\ast}$ is the temperature in that $\tau=\tau_{\min}$ for
%that value of $q$. For each $q$, the temperature range was covered by
%increments $\Delta T=0.05.$\ All curves behave very similarly around
%$\mathcal{T}_{q}=0$.}}{\Qlb{figure2}}{figure2.eps}%
%{\special{ language "Scientific Word";  type "GRAPHIC";
%maintain-aspect-ratio TRUE;  display "USEDEF";  valid_file "F";
%width 5.047in;  height 5.2217in;  depth 0pt;  original-width 8.0281in;
%original-height 8.3083in;  cropleft "0";  croptop "1";  cropright "1";
%cropbottom "0";  filename 'figure2.eps';file-properties "XNPEU";}}}%
%BeginExpansion
%EndExpansion

\bigskip%

%EndExpansion

\bigskip%

%TCIMACRO{\FRAME{ftbpFU}{5.047in}{1.6648in}{0pt}{\Qcb{Table 1 -- Characteristic
%folding (MC) time $t$ for three target (native) structures; the unit MC time
%used here corresponds to 8100 attempts to move the chain. For each structure
%and temperature $T_{i}$ there is a specific $q=q(T_{i})\geq$ 1 that minimizes
%$\tau$ (shaded cells). Due to the higher topological complexity of structure
%ID 36335, its $\tau$ is 5 to 10 times larger than the corresponding values of
%$\tau$ for the others two structures. A set of $N=150$ independent runs was
%used to estimate $\tau$ for each pair $(T,q)$. The figures were rounded off
%according to average relative uncertainty $\delta\tau=10\%$ (two significant
%figures); see text. At $T=1$, the values for $q$ that give the smallest
%folding characteristic times are generally depending on the complexity of the
%corresponding native structure --shaded cells, bold figures.}}{\Qlb{table1}%
%}{table1.eps}{\special{ language "Scientific Word";  type "GRAPHIC";
%maintain-aspect-ratio TRUE;  display "USEDEF";  valid_file "F";
%width 5.047in;  height 1.6648in;  depth 0pt;  original-width 8.2123in;
%original-height 2.6792in;  cropleft "0";  croptop "1";  cropright "1";
%cropbottom "0";  filename 'table1.eps';file-properties "XNPEU";}}}%
%BeginExpansion
\begin{figure}
[ptb]
\begin{center}
\includegraphics[
height=1.6648in,
width=5.047in
]%
{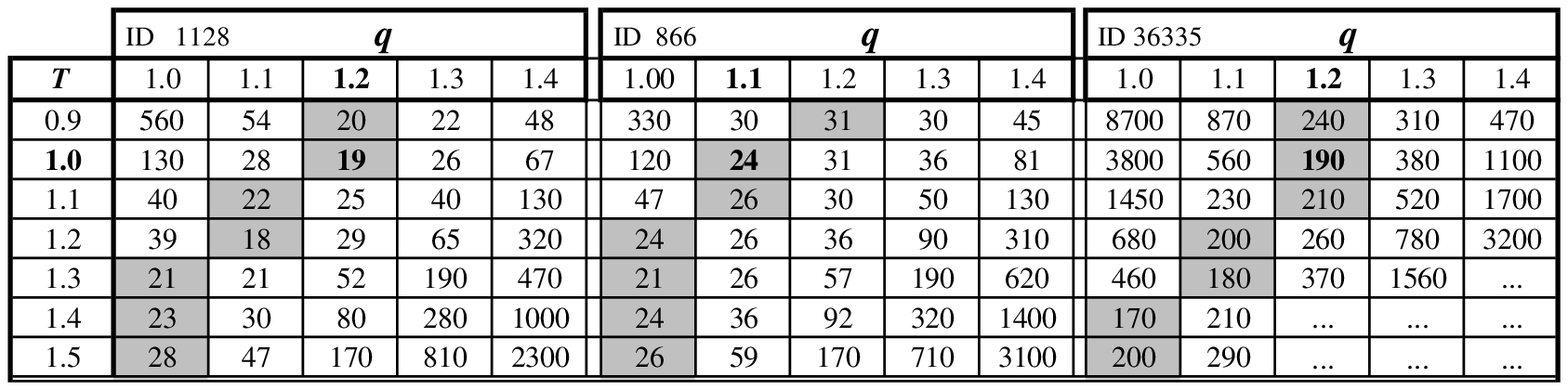}%
\caption{Table 1 -- Characteristic folding (MC) time $t$ for three target
(native) structures; the unit MC time used here corresponds to 8100 attempts
to move the chain. For each structure and temperature $T_{i}$ there is a
specific $q=q(T_{i})\geq$ 1 that minimizes $\tau$ (shaded cells). Due to the
higher topological complexity of structure ID 36335, its $\tau$ is 5 to 10
times larger than the corresponding values of $\tau$ for the others two
structures. A set of $N=150$ independent runs was used to estimate $\tau$ for
each pair $(T,q)$. The figures were rounded off according to average relative
uncertainty $\delta\tau=10\%$ (two significant figures); see text. At $T=1$,
the values for $q$ that give the smallest folding characteristic times are
generally depending on the complexity of the corresponding native structure
--shaded cells, bold figures.}%
\label{table1}%
\end{center}
\end{figure}
%EndExpansion

\end{document}